 \documentclass[12pt]{article}
 \newcommand{\be}{\begin{eqnarray}}
 \newcommand{\ee}{\end{eqnarray}}
 \newcommand{\nn}{\nonumber\\}
\begin{document}

\begin{center}
{\large\bf A Model Independent Approach to
 \\
Semi-Inclusive Deep Inelastic Scattering}~\footnote{ talk
 given by E. Christova at the 16th Int. Spin Physics Symposium, SPIN2004, Trieste, Italy}
\\
\vspace{.5cm}

{\large Ekaterina Christova$^\dagger$\footnote{Work supported by Grant Ph-1010 of the Bulgarian National Science Foundation},  Elliot Leader$^{\dagger\dagger}$}\\
{\it $\dagger$ Institute for Nuclear Research and Nuclear Energy,
Sofia,  echristo@inrne.bas.bg\\
$\dagger\dagger$ Imperial College,  London,  e.leader@imperial.ac.uk}
\end{center}
\vspace{0.1cm}

\begin{center}
{\bf Abstract}
\end{center}
We present a method for extraction of  detailed information 
on polarized quark densities 
 from semi-inclusive deep inelastic scattering $l+N\to l+h+X$,
 in both LO and NLO QCD without any assumptions about fragmentation
 functions and polarized sea densities. The only symmetries utilised are charge
 conjugation  and isotopic spin invariance of strong interactions.

\setcounter{page}{1}
\vspace{0.5cm}

{\bf 1.} 
At present the possibility to obtain a full information on
 the parton helicity densities 
 in a polarized nucleon 
 is related to semi-inclusive deep inelastic
 scattering (SIDIS) experiments of polarized leptons on polarized nucleons:
\begin{eqnarray}
 \overrightarrow{e} + \overrightarrow{N}\to e+ h+ X,\qquad h =\pi^\pm,\, K^\pm,...\label{SIDIS}
\ee
The first polarized SIDIS measurements were done by the 
SMC [1] and HERMES [2] collaborations,
 where the asymmetry $A_N^h$ was measured:
 \begin{eqnarray}
  A_{N}^h= \frac{1+(1-y)^2}{2y\,(2-y)}\,\frac{\Delta
\sigma^{h}_N}{\sigma^{h}_N}=
\frac{\sum e^2_q\,(\Delta q\, D_q^h + \Delta \bar q\, D_{\bar q}^h)
}{\sum e^2_q \,(q\,D_q^h + \bar q\,D_{\bar q}^h)},\qquad(LO)\label{A1}.
\end{eqnarray}
Here $\Delta\sigma^h_N$ ( $\sigma^h_N$) is the measured polarized 
(unpolarized) cross section.
Though this asymmetry allows, in principle,  a full extraction of 
the polarized parton densities, in practice, however, it faces two major
 problems. First, a good  knowledge of the fragmentation functions (FFs)
 $D_q^h$ is required. At present, these functions are rather poorly known
 and a number of model dependent assumptions are made.
 And second, it is not clear how to extend the ``purity'' method, used
 in the analysis in LO,  to the case of  NLO. 
 As the fulfilled and planned  measurements of polarized SIDIS 
 are done at rather 
 low $Q^2$, NLO corrections have to be taken into account.

\vspace{.2cm}

{\bf 2.} Here we show [3,4] that a measurement of the difference asymmetries $ A_{N}^{h^+ - h^-}$ and ratios 
$R_N^{h^+ - h^-}$:
\begin{eqnarray}
 A_{N}^{h^+ - h^-}=\frac{1+(1-y)^2}{2y\,(2-y)}\,\frac{\Delta\sigma_N^{h+} -
\Delta\sigma_N^{h-}} {\sigma_N^{h^+} - \sigma_N^{h^-}},\quad 
R_N^{h^+ - h^-}= \frac{\sigma_N^{h+} -
\sigma_N^{h-}}{\sigma_N^{DIS}},\label{Ah+h-}
 \end{eqnarray}
gives a lot of information on the polarized quark densities, in both LO and NLO, without any 
assumptions about the FFs. In general this can be traced as follows. C-invariance of strong 
interactions leads to  
$D_G^{h-\bar h}=0$ and $D_q^{h-\bar h}= -D_{\bar q}^{h-\bar h}$, which immediatelly  singles out
in  $ A_{N}^{h^+ - h^-}$ and  
$R_N^{h^+ - h^-}$  quantities that are  flavour non-singlets for both the parton
densities and the FFs. This implies that:
1) the badly known gluons  $G$, $\Delta G$ and $D_G^{h-\bar h}$ do not enter
 these asymmetries, 2) in its $Q^2$-evolution $ A_{N}^{h^+ - h^-}$ and  $ R_{N}^{h^+ - h^-}$ do
 not mix with other quantities, apart from those entering them
 and,  
3) $ A_{N}^{h^+ - h^-}$ can give
 information only on non-singlet parton densities: $\Delta u_V$, $ \Delta d_V$,
$\Delta\bar u -\Delta\bar d$, $\Delta s-\Delta\bar s$, etc, but in a model independent way. 
This holds in general and is true in any order in QCD.
 Depending on the SU(2) properties of the detected
 final hadron $h$, one obtains different pieces of information.
 This method for extracting the polarized quark densities will be used 
at Jefferson Lab.~[5], in the planned experiment E04-113, Semi-SANE.

\vspace{.2cm}

{\bf 3.} If $h=\pi^\pm$, SU(2) and C-invariance reduce the number of independent pion FFs: 
$D_u^{\pi^+-\pi^-} = - D_d^{\pi^+-\pi^-}, \quad D_s^{\pi^+-\pi^-} =0$. 
Then in LO we have:
\begin{eqnarray}
 A_{p}^{\pi^+-\pi^-}(x, z,Q^2) =\frac{4\Delta u_V
-\Delta d_V}{4u_V -d_V}(x,Q^2)\nn
 A_{n}^{\pi^+-\pi^-}(x, z,Q^2) =\frac{4\Delta d_V
  -\Delta u_V}{4d_V -u_V}(x,Q^2)\label{LO}
 \end{eqnarray}
Thus, the FFs cancel and $\Delta u_V$ and $\Delta d_V$ are 
expressed  solely in terms of measurable quantities and the known 
unpolarized $u_V$ and $d_V$. Note that the measurable
 quantities (  l.h.s.)  depend, in general, on $(x, z,Q^2)$, while the LO expressions ( r.h.s.) depend  on 
$(x,Q^2)$ only. This can serve as a test for the LO approximation
 or as an estimate of the theoretical systematic error when applying LO.

When going from LO to  NLO, the principle difference is that the simple products are
 replaced by convolutions
 and that gluons enter
 the cross sections. As explained above, in $ A_{N}^{\pi^+-\pi^-}$ the gluon 
contributions always drop out and we obtain:
\begin{eqnarray}
A_{p}^{\pi^+ -\pi^-} &=& \frac{ (4\Delta u_V -\Delta d_V)
[1+ \otimes (\alpha_s/2\pi) \Delta C_{qq}\otimes ] D_u^{\pi^+
-\pi^-}} {(4u_V -d_V)[1 + \otimes (\alpha_s /2\pi) 
C_{qq}\otimes ] D_u^{\pi^+ -\pi^-}}\nn
A_{n}^{\pi^+
-\pi^-}& =& \frac{ (4\Delta d_V -\Delta u_V) [1+ \otimes (\alpha_s
/2\pi) \Delta C_{qq}\otimes ]D_u^{\pi^+ -\pi^-}} {(4d_V
-u_V)[ 1+ \otimes (\alpha_s /2\pi) { C}_{qq}\otimes ]
D_u^{\pi^+ -\pi^-}}.\label{NLO}
\end{eqnarray}
The  FF  $D_u^{\pi^+-\pi^-}$, that enters (\ref{NLO}),  can be determined in
 unpolarized SIDIS:
\begin{eqnarray}
&&R_p^{\pi^+-\pi^-} = \frac{[4u_V -d_V][1+\otimes
(\alpha_s/2\pi){ C}_{qq}\otimes ] D_u^{\pi^+
-\pi^-}}{18F_1^p\,[1+2\gamma (y)\,R^{\,p}]}
\nn
&&R_n^{\pi^+-\pi^-} =\frac{[4d_V -u_V][1+\otimes (\alpha_s/2\pi)
{ C}_{qq}\otimes ]D_u^{\pi^+ -\pi^-}}{18F_1^n\,[1+2\gamma
(y)\,R^{\,n}]},\label{RNLO}
 \end{eqnarray}
where $\gamma (y) = (1-y)/[1+(1-y)^2]$. 

 Recently the HERMES collaboration published [6] very precise data for unpolarized SIDIS
 for $\pi^\pm$ production, that allows to determine
 $D^{\pi^+-\pi^-}_u$ directly,  without requiring knowledge of the other FFs:
\begin{eqnarray}
D^{\pi^+-\pi^-}_u = \frac{9\,(R_p^{\pi^+} - R_p^{\pi^-})\, \sigma^{DIS}_p}{4\,u_V -d_V}.
\end{eqnarray}
This was done in [7],  
 where $D_{u,d,s}^{\pi^+}$ were determined separately
 combining the HERMES  and the LEP $e^+e^-$ inclusive data on
 $\pi^\pm$-production. For the first time the $u$-quark FFs of the pions were obtained by the EMC collaboration [8].

\vspace{.2cm}

{\bf 4.} Having thus determined $\Delta u_V$ and $\Delta d_V$ we can proceed and 
 determine the SU(2) breaking of the polarized sea quarks. We have:
\begin{eqnarray}
(\Delta \bar u -\Delta\bar  d) =
\frac{1}{6}\left[\Delta q_3 +\Delta d_V -\Delta
u_V\right], \quad {\rm where}\quad \Delta q_3 = (\Delta u + \Delta\bar u) - (\Delta d-\Delta\bar d). 
\end{eqnarray}
In LO the valence quarks are deterimend via (\ref{LO}) and  $\Delta q_3$ is determined by $\Delta q_3= 
g_1^p(x,Q^2) -g_1^n(x,Q^2) $. In NLO  the valence quarks are deterimend via (\ref{NLO}) and 
 $\Delta q_3$ is obtained  by the NLO expression:
\begin{eqnarray}
g_1^p(x,Q^2) -g_1^n(x,Q^2) =\frac{1}{6} \Delta q_3\otimes \left(1+
\frac{\alpha_s(Q^2)}{2\pi}\delta C_q\right).
 \end{eqnarray}
 As SU(2) breaking is expected to be small,
 an  NLO treatment should be important.

Thus,  $A_N^{\pi^+-\pi^-}$ and  $g_1^N$  determine $\Delta u_V$, $\Delta d_V$ and $\Delta\bar u - \Delta\bar d$, 
, both in LO and NLO,
 without any assumptions. 
  Note that, due to $D_s^{\pi^+-\pi^-}=0$,
 even the commonly used 
 $s=\bar s$ or $\Delta s =\Delta\bar s$ are not made.

\vspace{.2cm}

{\bf 5.} If $h=K^\pm$ ( $K^0$ are not measured)    
 we cannot use isospin symmetry. However, if we make the 
natural assumption that unfavoured transitions  for $K^+$ and $K^-$ 
are equal (but not small): $D_d^{K^+-K^-}=0$, then 
 we can use $A_N^{K^+-K^-}$ to determine 
$(\Delta s - \Delta\bar s)\,D_s^{K^+-K^-}$. We assume that $\Delta u_V$ and $\Delta d_V$ 
are already determined from the pion SIDIS. Note that as the strange quark is a valence quark
 for $K^\pm$, $ D_s^{K^+ -K^-}$ is not small and $(\Delta s - \Delta\bar s)\,D_s^{K^+-K^-}$ would be nonzero 
only if  $(\Delta s - \Delta\bar s) \not= 0$.

 In LO we have:
 \begin{eqnarray}
A_{p}^{K^+ -K^-} &=&
\frac{4\Delta u_V D_u^{K^+ -K^-} +
(\Delta s-\Delta\bar s)D_s^{K^+ -K^-}} {4 u_V D_u^{K^+ -K^-}+ (s-\bar s) D_s^{K^+ -K^-} }
\nn
A_{n}^{K^+ -K^-}&=&
\frac{4\Delta d_V D_u^{K^+ -K^-} +
(\Delta s-\Delta\bar s)D_s^{K^+ -K^-}} {4 d_V D_u^{K^+ -K^-} +  (s-\bar s) D_s^{K^+ -K^-} }
\label{LOs}
\end{eqnarray}
We determine  $D_u^{K^+ -K^-}$ and
 $ (s-\bar s) D_s^{K^+ -K^-}$ 
 from unpolarized SIDIS:
\begin{eqnarray}
R_p^{K^+ -K^-} &=&\frac{4\, u_V\,D_u^{K^+ -K^-} + (s-\bar s) D_s^{K^+ -K^-}}{18 F_1^p}\nn
R_n^{K^+ -K^-} &=&\frac{4\, d_V\,D_u^{K^+ -K^-} + (s-\bar s) D_s^{K^+ -K^-}}{18\,F_1^n}.\label{RK}
\end{eqnarray}
Following the same argument as above, these ratios will shed light on the $(s-\bar s)$ difference. 

In NLO  the same quantities 
enter and  the expressions are analogous to (\ref{LOs})-(\ref{RK}), where
 the simple products are replaced by convolutions. 
The full expressions can be found in [4].
 Here we present the  considerably simpler formulae which result  when one takes $s=\bar s$:
\begin{eqnarray}
A_{p}^{K^+ - K^-} &=& \frac{[4\Delta u_V
 D_u^{K^+ - K^-} +
 (\Delta s - \Delta \bar s) D_s^{K^+ - K^-}]\,
 [1+\otimes  (\alpha_s/2\pi)  \Delta C_{qq}\otimes ] }
{4\, u_V\, [1+ \otimes\, (\alpha_s/2\pi )\, C_{qq}\otimes ]\, D_u^{K^+ - K^-}}\nn
A_{n}^{K^+ - K^-} &=& \frac{[4\,\Delta d_V  D_u^{K^+ - K^-} +
 (\Delta s - \Delta \bar s)D_s^{K^+ - K^-}]\,
 [1+\otimes  (\alpha_s/2\pi ) \Delta C_{qq}\otimes ]}
{4\, d_V\, [1+\otimes\,  (\alpha_s/2\pi )\, C_{qq}\otimes ]\, D_u^{K^+ - K^-}}
\end{eqnarray}
From unpolarized SIDIS we determine $D_u^{K^+-K^-}$:
\begin{eqnarray}
&&R_p^{K^+ - K^-} = \frac{2u_V\,[1+\otimes\,
(\alpha_s/2\pi){ C}_{qq}\,\otimes ]\, D_u^{K^+ - K^-}}{9F_1^p\,[1+2\gamma (y)\,R^{\,p}]}
\nn
&&R_n^{K^+ - K^-} =\frac{2d_V\,[1+\otimes\, (\alpha_s/2\pi)\,
{ C}_{qq}\,\otimes ]\, D_u^{K^+ - K^-}}{9F_1^n\,[1+2\gamma
(y)\,R^{\,n}]}.
 \end{eqnarray}

\vspace{.2cm}

{\bf 6.} If $h=\Lambda , \bar\Lambda$, then SU(2) invariance implies $
D_u^{\Lambda -\bar\Lambda} = D_d^{\Lambda -\bar\Lambda}$ 
and no conditions on $D_s^{\Lambda -\bar\Lambda}$. Thus, $A_N^{\Lambda -\bar\Lambda}$ 
will give information on $(\Delta s-\Delta\bar s)$ without any assumptions. 
In LO we have:
\begin{eqnarray}
A_{p}^{\Lambda-\bar\Lambda} &=&
\frac{(4\Delta u_V +\Delta d_V) D_u^{\Lambda-\bar\Lambda} +
(\Delta s-\Delta\bar s)D_s^{\Lambda-\bar\Lambda}} {(4 u_V+d_V) D_u^{\Lambda-\bar\Lambda}+ 
(s-\bar s) D_s^{\Lambda-\bar\Lambda} }
\nn
A_{n}^{\Lambda-\bar\Lambda}&=&
\frac{(4\Delta d_V + \Delta u_V) D_u^{\Lambda-\bar\Lambda} +
(\Delta s-\Delta\bar s)D_s^{\Lambda-\bar\Lambda}} {(4 d_V+u_V) D_u^{\Lambda-\bar\Lambda} + 
 (s-\bar s) D_s^{\Lambda-\bar\Lambda} }.
\end{eqnarray}
Unpolarized SIDIS  could  determine  $D_u^{\Lambda-\bar\Lambda}$ and 
$(s-\bar s) D_s^{\Lambda-\bar\Lambda}$:
\begin{eqnarray}
R_p^{\Lambda-\bar\Lambda} &=&\frac{(4\, u_V+d_V)D_u^{\Lambda-\bar\Lambda} +
 (s-\bar s) D_s^{\Lambda-\bar\Lambda}}{18 F_1^p}\nn
R_n^{\Lambda-\bar\Lambda} &=&\frac{(4\, d_V +u_V)D_u^{\Lambda-\bar\Lambda} + 
(s-\bar s) D_s^{\Lambda-\bar\Lambda}}{18\,F_1^n}.
\end{eqnarray}
 The expressions in NLO are straightforward and  can be found in [4].


\end{document}